\documentclass[aps,pra,reprint,floatfix]{revtex4-1}

\usepackage{amsmath}
\usepackage{physics}
\usepackage{graphicx}

\usepackage{natbib}
\begin{document}


\title{Ultrafast switching of persistent electron and hole currents in ring molecules}

\author{Tennesse Joyce}
\author{Agnieszka Jaron}

\affiliation{
JILA and Department of Physics, University of Colorado, Boulder, CO-80309, USA
}
\date{\today}

\begin{abstract}
A circularly polarized laser pulse can induce persistent intra-molecular currents by either exciting or ionizing molecules.
These two cases are identified as electron currents and hole currents, respectively, and up to now they have been studied only separately.
We report \textit{ab initio} time-dependent density-functional theory (TDDFT) simulations of currents during resonance-enhanced two-photon ionization of benzene, which reveal for the first time that both electron and hole currents can be present simultaneously. 
By adjusting the intensity of the laser pulse, the balance between the two types of current can be controlled, and the overall sign of the current can be switched.
We provide a physical explanation for the effect in terms of complex molecular orbitals which is consistent with the TDDFT simulations.
\end{abstract}

\maketitle

It has long been understood that, in response to an applied magnetic field, the delocalized electrons of an aromatic molecule circulate in  so-called aromatic ring current \cite{gomes_aromaticity_2001,krygowski_aromaticity_2014}. This
effect is important in nuclear magnetic resonance spectroscopy, where the internal magnetic
field generated by the ring current is responsible for diamagnetic shielding \cite{heine_magnetic_2005}.
In 2006, it was proposed that ring currents in molecules could also be induced by ultrashort laser pulses with circular or elliptical polarization \cite{barth_periodic_2006,barth_unidirectional_2006}.
The basic mechanism is that angular momentum carried by light is transfered to
electrons in a molecule.
Due to conservation of angular momentum, the current persists after the pulse has ended---even without an external magnetic field.
Various experiments on atomic targets have confirmed the existence of the effect \cite{wollenhaupt_photoelectron_2009,eckart_ultrafast_2018}, although no direct observational data is available in the case of molecules.
Recent interest in photoinduced ring currents is motivated by the rapid technological advances in polarization control of high-harmonic radiation made in the last few years \cite{fleischer_spin_2014,hickstein_non-collinear_2015,huang_polarization_2018}, which may enable experimental study of these phenomena in the near future \cite{neufeld_background-free_2019}.

There are several major
advantages of photoinduced ring currents compared to those induced by static magnetic fields.
First, the current is expected to be orders
of magnitude stronger, and so is the  induced  magnetic field \cite{yuan_attosecond-magnetic-field-pulse_2013}.
Second, they enable femtosecond
(or even attosecond) time-resolved studies of aromaticity and magnetism \cite{ulusoy_correlated_2011,yuan_attosecond_2017}.
Lastly, they
establish the possibility for coherent control of ring currents \cite{mineo_quantum_2017}, which may have applications for
controlling chemical reactions or the operation of advanced opto-electronic devices.

In this Letter we predict a novel effect which causes the dominant charge carrier of the ring current to transition from electrons to holes as the peak laser intensity increases past around $10^{12}$ W/cm$^2$.
We illustrate the effect with a series of \textit{ab initio} time-dependent density functional theory (TDDFT) simulations of benzene (C$_6$H$_6$), which is the prototypical aromatic molecule.
Lastly, we demonstrate that the effect is not accounted for in the commonly used few level model of ring currents, due to the fact that it neglects ionization.
This calls into question the results of several previous studies (e.g. \cite{barth_periodic_2006,barth_unidirectional_2006,mineo_quantum_2017}) where it was assumed that the few level model is accurate for laser intensities on the order of $10^{12}$ W/cm$^2$.

We begin by introducing the distinction between electron and hole current: when an electron is promoted to an orbital with nonzero angular momentum, this creates an electron current;
when an electron is removed (e.g., ionized) from an orbital with nonzero angular momentum, this creates a hole current.
So far, hole currents have mostly been studied in the context of strong field ionization of atoms by circularly polarized laser pulses, and it was recently confirmed experimentally that a hole can be created with a specific angular momentum relative to the laser polarization \cite{barth_nonadiabatic_2011,herath_strong-field_2012,zhu_helicity_2016,liu_nonadiabatic_2016}.
Electron currents on the other hand do not involve ionization, only excitation.

However, in the interaction of atoms and molecules with strong laser fields, excitation and ionization are often closely related and occur together.
A typical example is resonance-enhanced multiphoton ionization (REMPI)
\cite{antonov_stepwise_1978,dietz_model_1982}, a two-step ionization process wherein an atom or molecule is first excited to an intermediate state (that must be resonant with some multiple of the laser frequency) and then subsequently ionized.
Now consider REMPI in a system where the intermediate excited state corresponds to an electron current, and the final ionized state corresponds to a hole current (we will show that benzene is such a system).
The balance between excitation and ionization (and therefore electron and hole current) will depend on the laser intensity because the processes involve different numbers of photons (and therefore scale with different powers of intensity).
In particular at low intensities we expect electron current to dominate (excitation), and at high intensities we expect hole current to dominate (ionization).

\begin{figure*}[t]
\includegraphics[width=0.7\linewidth]{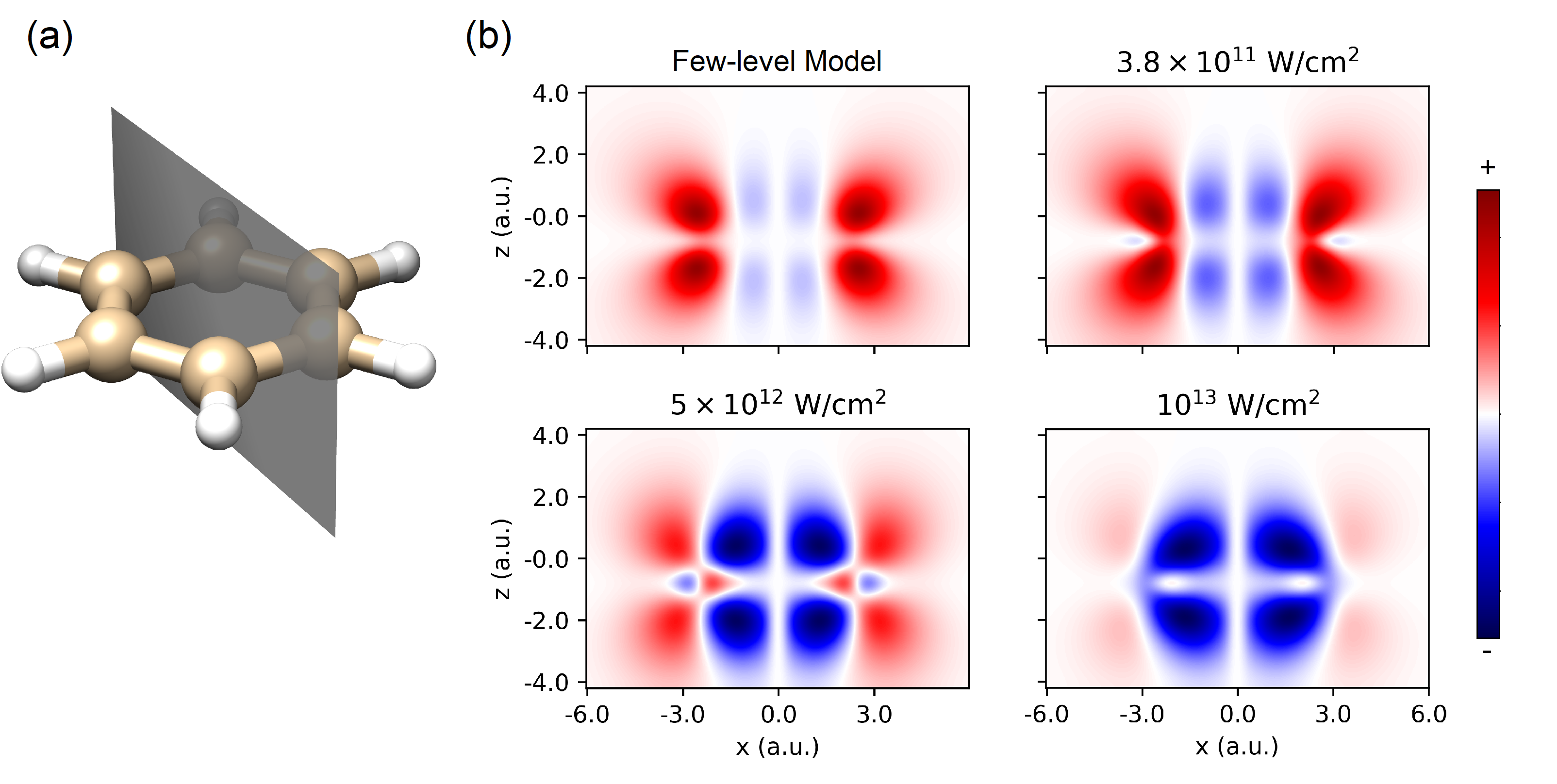}
    \caption{
    (a) 
    Visualization of
    the current density based on the component passing through a plane bisecting the molecule as shown (averagea over all possible orientations of that plane [see Eq. (\ref{eq:jphi})]) 
    (b) Cross sections of the current density taken at the end of the laser pulse ($t=200$ a.u.) for several different simulations.
    At low laser intensity the co-rotating current (red) dominates, while at high intensity the counter-rotating current (blue) dominates.
Note:     
    Each plot is scaled individually relative to the maximum absolute value within that plot.
    The nuclei lie in the plane $z=0$ with the carbon ring at $x=\pm 2.63$ a.u. and the hydrogen ring at $x=\pm 4.69$ a.u..
    }
\label{fig:shape}
\end{figure*}

Our main theoretical method is TDDFT, as implemented by Octopus \cite{marques_octopus:_2003,castro_octopus:_2006,andrade_real-space_2015}, which provides a fully nonperturbative description of the light-matter interaction.
As a reference point to compare against the full TDDFT simulations, we also consider the few level model of ring currents (e.g. \cite{barth_unidirectional_2006}).
We discuss the implementations of both models in \footnote{See Supplemental Material for details on the TDDFT and few-level models, orientation dependence, and the method for interpolating over intensity.}.
Because the few level model does not include ionization, we expect the two models to diverge at high enough laser intensities.

The laser pulse in our simulations is described in the dipole approximation by the following electric field,
\begin{align}
\label{eq:pulse}
\boldsymbol{\mathcal{E}}(t) =
\begin{cases}
\mathcal{E} \sin^2\left(\pi t/T\right)\text{Re}\left[\hat{\epsilon}  e^{i\omega (t-T/2)}\right], & 0<t<T,\\
0,   &   \text{otherwise},
\end{cases}
\end{align}
with central frequency $\omega=6.76$ eV (183 nm),
duration $T = 16\pi/\omega =202 ~\text{a.u.} = 4.9$ fs, circular polarization $\hat{\epsilon}=(\hat{x}+i\hat{y})/\sqrt{2}$ (with the molecule in the $xy$-plane),
and a variable peak amplitude $\mathcal{E}$.
The central frequency was chosen to be resonant with the doubly degenerate E$_{1u}$ state (as computed with linear response TDDFT \cite{Note1}), which is predominantly associated with the HOMO-LUMO transition (HOMO = Highest Occupied Molecular Orbital; LUMO = Lowest Unoccupied Molecular Orbital). Because the computed ionization threshold is $9.0$ eV $<2\omega$, this laser pulse is designed to drive 1+1 REMPI where one photon is enough to promote electron to the excited state and one additional photon to ionize.

After interacting with the laser pulse ($t>T$), the benzene molecule is in a superposition of the A$_{1g}$ ground state and the E$_{1u}$ excited state and also, to an extent, ionized.
This causes oscillations in the charge and current densities $\rho(\mathbf{r},t)$ and $\mathbf{J}(\mathbf{r},t)$, respectively,  with period 612 as (corresponding to the energy difference between the ground state and excited states), which are an example of attosecond charge migration \cite{worner_charge_2017}.

\begin{figure}[t]
\includegraphics[width=\linewidth]{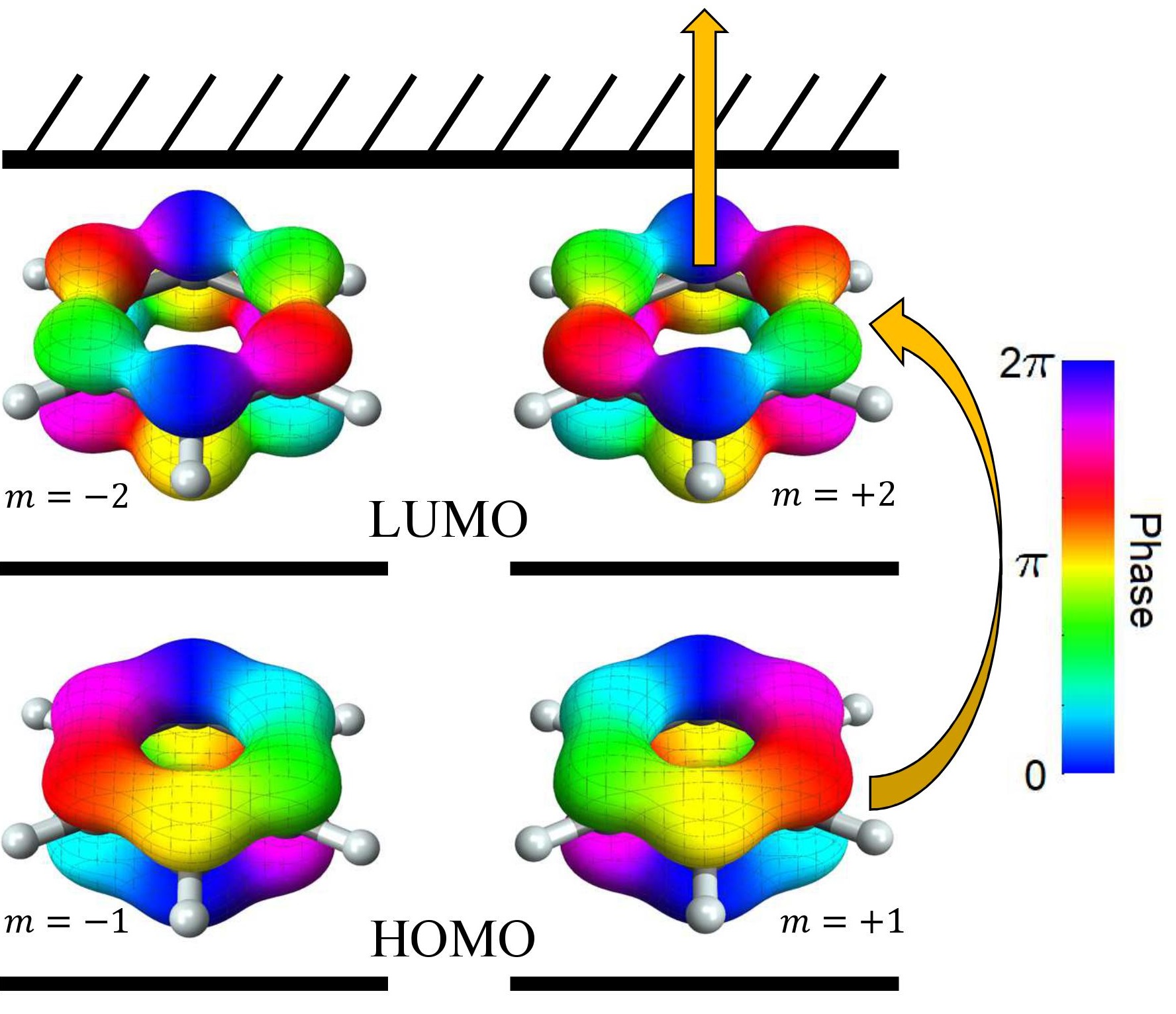}
    \caption{
    Schematic illustrating the complex molecular orbitals and the physical mechanism for the transition from electron to hole current.
    Color indicates the complex phase.
    }
\label{fig:schematic}
\end{figure}

In order to visualize the current we isolate the stationary component of the current density, by computing an angle averaged cross section defined by the following integral (in cylindrical coordinates $\rho,z,\phi$),
\begin{align}
\label{eq:jphi}
    J(x,z,t) = \frac{1}{2\pi} \int_{0}^{2\pi} \hat{\phi}\cdot\mathbf{J}(|x|,z,\phi)d\phi.
\end{align}
The geometric interpretation of this integral is given in Fig. \ref{fig:shape}.  {  The angle averaging procedure for the few level model causes that the fast-oscillating component effectively vanishes}.
Within the few-level model, the fast-oscillating component of the current density is zeroed out by this averaging procedure because of its parity.
{ It has similar effect on TDDFT results, and therefore $J(x,z,y)$ has only a very gradual time dependence for $t>T$.} The same is true for TDDFT.
These integrated current densities are plotted in Fig. \ref{fig:shape}b.
At low intensities the current is a combination of a strong co-rotating current (red) and a weak counter-rotating current (blue), while at high intensities the counter-rotating current dominates.
As we explain below (see Fig. \ref{fig:schematic}), the reversal is a signature of the transition from electron to hole current regime.

\begin{figure*}[t]
\includegraphics[width=\linewidth]{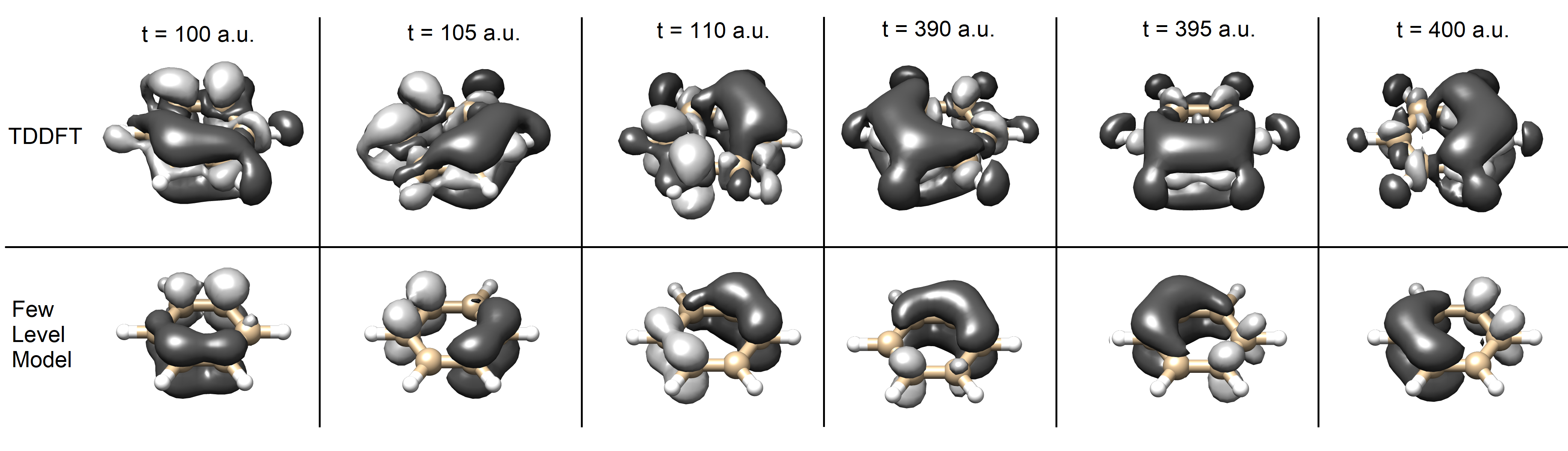}
    \caption{
     Snapshots of the charge displacement
     induced by a circularly-polarized laser pulse with peak intensity
     $5\times 10^{12}$ W/cm$^2$
     taken around the peak of the laser pulse (first three columns $t\approx 100$ a.u.) and  after the laser pulse (last three columns $t\approx 400$ a.u.). 
     Light areas indicate excess electrons while dark areas indicate fewer electrons, as compared to the ground state charge density before the laser pulse.
     We compare the results between the two theoretical models, TDDFT (top row) and the few level model (bottom row).
    }
\label{fig:hole_density}
\end{figure*}
\begin{figure}[t]
\includegraphics[width=\linewidth]{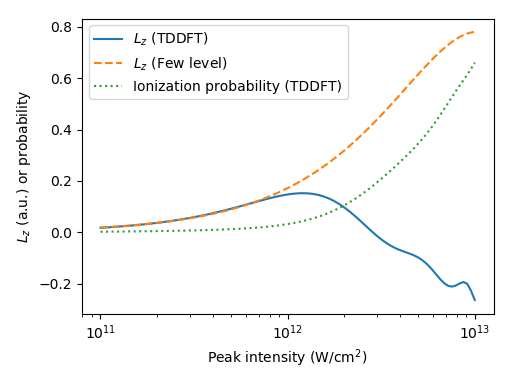}
    \caption{
    Comparison of full TDDFT simulations (solid blue line) to the few level model (orange dashed line).
    For peak intensities, when ionization (dotted green line) becomes non-negligible, the two models begin to disagree.
    The smooth lines have been interpolated between the calculated intensities using the method described in \cite{Note1}.
    }
\label{fig:comparison}
\end{figure}

The oscillatory component of the charge motion is best visualized by plotting the charge displacement,
\begin{align}
    \Delta\rho(\mathbf{r},t) = \rho(\mathbf{r},t) - \rho(\mathbf{r},0),
\end{align}
shown in Fig. \ref{fig:hole_density}.
The cloud of displaced charge circulates around the molecule with the expected period of 612 as, and this continues even after the pulse ends.
Overall, both the magnitude and shape of the charge displacement are remarkably similar between the two models, however there are some subtle differences.
First, long after the laser pulse the two models gradually become desynchronized.
Second, in TDDFT there appears to be a rearrangement of charge in the plane of the molecule, whereas the few level model only predicts the dynamics above and below the plane.

Another important observation about the density difference is that the dark areas are generally larger than the light areas.
In the TDDFT results one reason for this is ionization, with the ionization probability given by
\begin{align}
\label{eq:ionization_def}
    P^\text{ionize} = -\int 
    \Delta\rho(\mathbf{r},2T)d^3\mathbf{r},
\end{align}
where the integral ranges over the simulation box. 
Unexpectedly, the few level model also appears to have dark areas larger than light areas even though it does not include ionization, and in fact the charge displacement must integrate to zero in that model.
The reason for this is that the E$_{1u}$ is of mixed character, part of which involves excitation to LUMO + 3
\cite{Note1}.
{ Note: The excess of darker areas in the TDDFT model is a combination of both ionization and excitation to LUMO +3 orbital}.

{ The intensity dependence of the dynamics is illustrated in Fig. \ref{fig:comparison}.
using  the current. Note: this current is directly proportional to $z$-component of the magnetic moment as well as  $z$-component of electronic angular momentum),}
Since the domain of integration is the simulation box, ionized electrons are not included.
For this reason we plot $L_z(2T)$ so that the ionizing wavepacket has enough time to leave the box.
Whereas in the few level model the magnetic moment increases monotonically with the laser intensity (up to about $10^{13}$ W/cm$^2$, after which the system Rabi oscillates back to the ground state), in TDDFT the current starts to decrease already around $10^{12}$ W/cm$^2$, and reverses sign for even higher intensities. 
We also plot the ionization probability (defined in Eq. \ref{eq:ionization_def}), and conclude that the reversal occurs precisely when the ionization probability becomes non-negligible.

{ The implications of the transition from electron to hole current on the charge dynamics, and the
underlying physical mechanism responsible for that transition, can be  understood in more detail describe in more detail using complex molecular orbitals, as illustrated schematically in Fig. \ref{fig:schematic}.
These orbitals represent } a change of basis from the usual real-valued Kohn-Sham orbitals $\psi_n(\mathbf{r})$ (defined in \cite{Note1}),
\begin{align}
    \psi^\text{HOMO}_{\pm}(\mathbf{r}) = \left[\psi_{14}(\mathbf{r})\pm i\psi_{15}(\mathbf{r})\right]/\sqrt{2},\\
    \psi^\text{LUMO}_{\pm}(\mathbf{r}) = \left[\psi_{16}(\mathbf{r})\pm i\psi_{17}(\mathbf{r})\right]/\sqrt{2}.
\end{align}
The advantage of using complex orbitals is that they are eigenfunctions of the 6-fold symmetry operator (rotation about the molecular axis by 60$^\circ$),
\begin{eqnarray}
    \exp\left[-\frac{i\pi}{3 \hbar}\frac{\hat{L}_z}{\hbar}\right]\psi^\text{HOMO}_{\pm}(\mathbf{r}) = \exp\left[\mp\frac{i\pi}{3}\right]\psi^\text{HOMO}_{\pm}(\mathbf{r}),\\
    \exp\left[-\frac{i\pi}{3}\frac{\hat{L}_z}{\hbar}\right]\psi^\text{LUMO}_{\pm}(\mathbf{r}) = \exp\left[\mp\frac{2i\pi}{3}\right]\psi^\text{LUMO}_{\pm}(\mathbf{r}).
\end{eqnarray}

The complex orbitals have magnetic quantum numbers $m$ defined modulo 6: $\psi^\text{HOMO}_{\pm}$ have $m=\pm 1$ and $\psi^\text{LUMO}_{\pm}$ have $m=\pm 2$.
Just as for atomic orbitals, the sign of $m$ indicates the direction the electron circulates around the molecule, and the magnitude indicates more-or-less the angular speed.
We have chosen our conventions such that $m>0$ electrons are co-rotating with the laser field, and $m<0$ electrons are counter-rotating.

{ Using the notation of complex orbitals,  Fig. \ref{fig:schematic} illustrates how
in the ground state}, both $\psi^\text{HOMO}_{\pm}$ are doubly occupied, and consequently there is zero net current.
When the benzene molecule is exposed to a circularly polarized laser pulse, the usual selection rule $\Delta m = 1$ applies (here we assume the laser is polarized in the molecular plane, see \cite{Note1} for the more general case), so that the only dipole-allowed transition is $\psi^\text{HOMO}_{+}$ to  $\psi^\text{LUMO}_{+}$, which is the dominant component of the E$_{1u}$ excited state.
The electron excited to LUMO contributes a strong co-rotating current ($m=+2$), but the imbalance of electrons in the HOMO contributes a weaker counter-rotating current ($m=-1$).
This can alternatively be interpreted as a positively charged hole occupying $\psi^\text{HOMO}_{+}$ producing a co-rotating hole current (rather than a counter-rotating electron current).
This is precisely what we see in the top row of Fig. \ref{fig:shape}b, two components to the current with opposite sign (red and blue).

In order to explain the reversal of the current at higher intensity (bottom row of Fig. \ref{fig:shape}b), we simply recognize that the electron previously excited to $\psi^\text{LUMO}_{+}$ can absorb a second photon from the same laser pulse, ionizing, and leaving behind only the hole current.
The balance between the one-photon excitation and the two-photon ionization processes can be controlled by varying the laser intensity, because the first process scales with $I$ while the second process scales with $I^2$ (with $I\propto \mathcal{E}^2$ the laser intensity).
Furthermore, it is now apparent that the sign reversal can be interpreted as a change in the dominant charge carrier from electrons to holes.

{ In conclusion, we have shown that both electron and hole currents are present during resonance-enhanced two-photon ionization of benzene, and the balance between the two current regimes can be controlled by varying the peak laser intensity.}
We have proposed a simple explanation for the effect in terms of molecular orbitals, which is consistent with the results of full TDDFT simulations.
Variants of complex orbital model should apply to a wide variety of molecules other than benzene, meaning that the structure of the complex molecular orbitals can be used to predict the interplay between electron and hole currents during REMPI.
In order to measure this effect in experiment, several pump-probe schemes have been proposed that are sensitive to the magnitude and direction of the ring current \cite{eckart_ultrafast_2018,neufeld_background-free_2019}. In [25], we demonstrate that the reversal is independent of the orientation of the molecule, which greatly simplifies any potential experiment.
Finally, our results suggest that the few level model typically used to study photoinduced ring currents may be insufficient even for moderate laser intensities around 10$^{12}$ W/cm$^2$.
A more \textit{ab initio} nonperturbative theory such as TDDFT, as used in present paper, is more appropriate for this regime.

This work was supported by the NSF Grant No. PHY-1734006 and Grant No.
PHY-2110628.
This work utilized resources from the University of Colorado Boulder Research Computing Group, which is supported by the National Science Foundation.

\bibliography{refs}

\end{document}